\documentclass[12pt]{article}
\usepackage{epsfig}
\def\beq{\begin{equation}}
\def\eeq#1{\label{#1}\end{equation}}
\def\eeqn{\end{equation}}

\def\beqa{\begin{eqnarray}}
\def\eeqa#1{\label{#1}\end{eqnarray}}
\def\eeqan{\end{eqnarray}}

\let\bar=\overbar

\def\Dslash{\not{\hbox{\kern-4pt $D$}}}
\def\dslash{\not{\hbox{\kern-2pt $\del$}}}

\def\msb{{\bar{\ssstyle M \kern -1pt S}}}

\usepackage{fancyhdr,graphicx}
\def\Title#1{\begin{center} {\Large {\bf #1} } \end{center}}

\begin{document}

\Title{Quark deconfinement phase transition in neutron stars}

\bigskip\bigskip


\begin{raggedright}

{\it Grigor Alaverdyan\index{Alaverdyan, G.}\\
Department of Radiophysics\\
Chair of Wave Processes Theory\\
Yerevan State University\\
Yerevan 0025\\
Armenia\\
{\tt Email: galaverdyan@ysu.am}}
\bigskip\bigskip
\end{raggedright}

 \begin{quote}
{\small{ \textbf{Abstract}

\textit{The hadron-quark phase transition in the interior of
compact stars is investigated, when the transition proceeds
through a mixed phase. The hadronic phase is described in the
framework of relativistic mean-field theory, when also the
scalar-isovector delta-meson mean-field is taken into account. The
changes of the parameters of phase transition caused by the
presence of delta-meson field are explored. The results of
calculation of structure of the mixed phase (Glendenning
construction) are compared with the results of usual first-order
phase transition (Maxwell construction).}

\textbf{Key words:} neutron stars: quarks: phase transition: mixed phase\\
\textbf{PACS numbers:} 97.60.Jd, 26.60.+c, 12.39.Ba }}
\end{quote}
\bigskip
\section{Introduction}
The structure of compact stars functionally depends on the
equation of state (EOS) of matter in a sufficiently wide range of
densities - from $7.9$ g/cm$^3$ (the endpoint of thermonuclear
burning) to one order of magnitude higher than nuclear saturation
density. Therefore, the study of properties and composition of the
matter constituents at extremely high density region is of a great
interest in both nuclear and neutron star physics. The
relativistic mean-field (RMF) theory \cite{Wal74,SW86,SW97} has
been effectively applied to describe the structure of finite
nuclei\cite{Lal97,Typel99}, the features of heavy-ion
collisions\cite{Ko_Li96,PR_Lal07}, and the equation of state (EOS)
of nuclear matter\cite{Mill95}. Inclusion of the scalar-isovector
$\delta$-meson in this theoretical scheme and investigation of its
influence on low density asymmetric nuclear matter was realized in
Refs.\cite{Kubis97,Liu02,Greco03}. At sufficiently high density,
different exotic degrees of freedom, such as pion and kaon
condensates, also deconfined quarks, may appear in the strongly
interacting matter. The modern concept of hadron-quark phase
transition is based on the feature of that transition, that is the
presence of two conserved quantities in this transition: baryon
number and electric charge\cite{Gl92}. It is known that, depending
on the value of surface tension, $\sigma_{s}$, the phase
transition of nuclear matter into quark matter can occur in two
scenarios \cite{Heis93}: ordinary first order phase transition
with a density jump (Maxwell construction), or formation of a
mixed hadron-quark matter with a continuous variation of pressure
and density (Glendenning construction)\cite{Gl92}. Uncertainty of
the surface tension values does not allow to determine the phase
transition scenario, taking place in realty. In our recent paper
\cite {Al09a} in the assumption that the transition to quark
matter is a usual first-order phase transition, described by
Maxwell construction, we have shown that the presence of the
$\delta$-meson field leads to the decrease of transition pressure
$P_{0}$, of baryon number densities $n_{N}$ and $n_{Q}$. In this
article we investigate the hadron-quark phase transition of
neutron star matter, when the transition proceeds through a mixed
phase. The results of calculation of structure of the mixed
phase(Glendenning construction) are compared with the results of
usual first-order phase transition (Maxwell construction).
Influence of $\delta$-meson field on phase transition
characteristics is discussed. Finally, using the EOS obtained, we
calculate the integral and structural characteristics of Neutron
stars with quark degrees of freedom.

\section{Neutron star matter equation of state}

\subsection{Nuclear matter}

In this section we consider the EOS of matter in the region of
nuclear and supranuclear density($n \geq 0.1$ fm$^{-3}$). For the
lower density region, corresponding to the outer and inner crust
of the star,  we have used the EOS of Baym-Bethe-Pethick
(BBP)\cite{BBP71}. For description of hadronic phase we use the
relativistic Lagrangian  density of many-particle system consisted
of nucleons, $p$, $n$, and  isoscalar-scalar ($\sigma$),
isoscalar-vector ($\omega$), isovector-scalar ($\delta$), and
isovector-vector ($\rho$) - exchanged mesons :
\begin{eqnarray}{\cal L}_{\sigma\omega\rho\delta}
(\sigma(x),\omega _{_{\mu }}(x),\vec{\rho }_{_{\mu }}(x),
\vec{\delta}(x)) ={ \cal L}_{\sigma\omega \rho}(\sigma(x),\omega
_{_{\mu }}(x),\vec{\rho }_{_{\mu }}(x))-U(\sigma(x))+ {\cal
L}_{\delta}(\vec{\delta}(x)),
\end{eqnarray}where $\cal L_{\sigma\omega \rho}$ is the linear part of
relativistic Lagrangian density without $\delta$-meson field
\cite{Gl00}, $U(\sigma)=\frac{b}{3}m_{N}\left( g_{\sigma }\sigma
\right) ^{3}+\frac{c}{4}\left( g_{\sigma }\sigma \right) ^{4}$ is
the $\sigma$-meson self-interaction term and

\begin{equation}\label{d}
{\cal L}_{\delta}(\vec{\delta})=g_{\delta } \bar {\psi}_{N}
\vec{\tau }_{N} \vec{\delta }\psi_ {N}+\frac{1}{2}(\partial _{\mu
}\vec{\delta}\partial ^{\mu}
\vec{\delta}-m_{\delta}\vec{\delta}^{2})
\end{equation}

\noindent  is the contribution of the $\delta$-meson field. This
Lagrangian density (1) contains the meson-nucleon coupling
constants, $g_{\sigma },~ g_{\omega },~g_{\rho },~g_{\delta}$ and
also parameters of $\sigma$-field self-interacting terms, $b$ and
$c$. In our calculations we take for the $\delta$ coupling
constant, $a_{\delta}=\left(g_{\delta}/m_{\delta}\right)^2=2.5$
fm$^2$, as in \cite{Liu02,Greco03,Al09a}, for the bare nucleon
mass, $m_{N}=938.93$ MeV, for the nucleon effective mass,
$m_{N}^{\ast}=0.78~m_{N}$, for the baryon number density at
saturation, $n_{0}=0.153$ fm$^{-3}$, for the binding energy per
baryon, $f_{0}=-16.3$ MeV, for the incompressibility modulus,
$K=300$ MeV, and for the asymmetry energy, $E_{sym}^{(0)}=32.5$
MeV. Five other constants
$a_{\sigma}=\left(g_{\sigma}/m_{\sigma}\right)^2$,
$a_{\omega}=\left(g_{\omega}/m_{\omega}\right)^2$,
$a_{\rho}=\left(g_{\rho}/m_{\rho}\right)^2$, $b$ and $c$ then can
be numerically determined\cite{Al09a}.
\begin{table}[h]
\begin{center}
\begin{tabular}{l|c|c|c|c|c|c}   & $a_{\sigma }$ (\ fm$^{2}$) & $a_{\omega }$ (\ fm$^{2}$) & $a_{\delta }$ (\ fm$^{2}$) &
$a_{\rho }$ (\ fm$^{2}$) & $b$ (\ fm$^{-1}$) & $c$   \\
\hline $\sigma \omega \rho $ & $9.154$ & $4.828$ & $0$ & $4.794$ &
$1.654\cdot 10^{-2}$ & $1.319\cdot 10^{-2}$ \\ \hline
$\sigma \omega \rho \delta $ & $9.154$ & $4.828$ & $2.5$ & $13.621$ & $%
1.654\cdot 10^{-2}$ & $1.319\cdot 10^{-2}$ \\ \hline
\end{tabular}
\caption{Model parameters without ($\sigma \omega \rho $) and with
($\sigma \omega \rho \delta$) a $\delta$ -meson field.}
\label{T-1}
\end{center}
\end{table}
In Table 1 we list the values of the model parameters with
($\sigma\omega\rho\delta$) and without ($\sigma\omega\rho$) the
isovector-scalar $\delta$ meson interaction channel.

The knowledge of the model parameters makes it possible to solve
the set of four equations in a self-consistent way and to
determine the re-denoted mean-fields, $\sigma \equiv
g_{\sigma}\bar {\sigma}$, $\omega \equiv g_{\omega}\bar
{\omega_{0}}$, $\delta \equiv g_{\delta}\bar{\delta}^{(3)}$, and
$\rho \equiv g_{\rho}\bar {\rho_{0}}^{(3)}$, depending on baryon
number density $n$ and asymmetry parameter $\alpha=(n_n-n_p)/n$.
The standard QHD procedure allows to obtain expressions for energy
density $\varepsilon(n,\alpha)$ and pressure $P(n,\alpha)$ of
nuclear $npe$ plasma:

\begin{eqnarray}{}
\varepsilon_{NM} ({n,\alpha},\mu_{e}) =
\frac{{1}}{{\pi^{2}}}\int\limits_{0}^{k_{-} (n,\alpha)} {\sqrt
{k^{2} + m_{p}^{\ast}( {\sigma,\delta})^{2}}}~ k^{2}dk +
\frac{{1}}{{\pi ^{2}}}\int\limits_{0}^{k_{+} (n,\alpha)} {\sqrt
{k^{2} + m_{n}^{\ast}( {\sigma,\delta})^{2}}}~ k^{2}dk
\nonumber\\+ \frac{{b}}{{3}}\,m_{N} \,\sigma ^{3} +
\frac{{c}}{{4}}\,\sigma ^{4} + \frac{{1}}{{2}}\left(
{\,\frac{{\sigma ^{\,2}}}{{a_{\sigma} } } + \frac{{\omega
^{2}}}{{a_{\omega} } } + \frac{{\delta ^{\,2}}}{{a_{\delta} } } +
\frac{{\rho ^{\,2}}}{{a_{\rho} }
}}\right)\nonumber\\+\frac{{1}}{{\pi^{2}}}\int\limits_{0}^{\sqrt{\mu_{e}^2-m_{e}^2}}
{\sqrt {k^{2} + {m_{e}}^{2}}}~ k^{2}dk,
\end{eqnarray}

\begin{eqnarray}{}
 P_{NM} ({n,\alpha},\mu_{e}) = \frac{1}{\pi
^{2}}\int\limits_{0}^{k_{-} (n,\alpha)} {\left( \sqrt {k_{-}
(n,\alpha)^2 + m_{p}^{\ast}({\sigma,\delta})^{2}}- \sqrt {k^{2} + m_{p}^{\ast}( {\sigma,\delta})^{2}}\right)\;}k^{2}dk  \nonumber \\
+\frac{{1}}{{\pi ^{2}}}\int\limits_{0}^{k_{+} (n,\alpha)}
{\left( \sqrt {k_{+} (n,\alpha)^2 + m_{n}^{\ast}({\sigma,\delta})^{2}}- \sqrt {k^{2} + m_{n}^{\ast}( {\sigma,\delta})^{2}}\right)}k^{2}dk \nonumber \\
- \frac{{b}}{{3}}\,m_{N} \,\sigma ^{3} - \frac{{c}}{{4}}\,\sigma
^{4}+ \frac{{1}}{{2}}\left( { - \frac{{\sigma ^{2}}}{{a_{\sigma} }
} + \,\frac{{\omega ^{2}}}{{a_{\omega }} } - \frac{{\delta
^{2}}}{{a_{\delta} } } + \frac{{\rho
^{2}}}{{a_{\rho}}}}\right)\nonumber\\+ \frac{{1}}{{3\pi ^{2}}}\mu
_{e} \left( {\mu _{e} ^{2} - m_{e}
^{2}}\right)^{3/2}-\frac{{1}}{{\pi^{2}}}\int\limits_{0}^{\sqrt{\mu_{e}^2-m_{e}^2}}
{\sqrt {k^{2} + {m_{e}}^{2}}}~ k^{2}dk,
\end{eqnarray}

\noindent where $\mu_{e}$ is the chemical potential of electrons,
\begin{equation}
m_{p}^{\ast}({\sigma,\delta})=m_{N}-\sigma-\delta,~~~m_{n}^{\ast}({\sigma,\delta})=m_{N}-\sigma+\delta
\end{equation}
\noindent are the effective masses of the proton and neutron,
respectively, and
\begin{equation}
k_{\pm}(n,\alpha)=
\left(\frac{{3\pi^{2}n}}{{2}}(1\pm\alpha)\right)^{1/3}.
\end{equation}

The chemical potentials of the proton and neutron are given by
\begin{eqnarray}
\label{} \mu_{p}(n,\alpha)=\sqrt {k_{F} \left( {n}
\right)^{2}\left( {1 - \alpha} \right)^{2/3} + \left( {m_{N} -
\sigma - \delta}
\right)^{\,2}} + \omega + \frac{1}{2}\rho,\\
\mu_{n}(n,\alpha)=\sqrt {k_{F} \left( {n} \right)^{2}\left( {1 +
\alpha} \right)^{2/3} + \left( {m_{N} - \sigma + \delta}
\right)^{\,2}}+ \omega - \frac{1}{2}\rho.
\end{eqnarray}

\begin{figure}[ht]
\begin{center}
  \begin{minipage}{0.47\linewidth}
   \begin{center}
   \epsfig{file=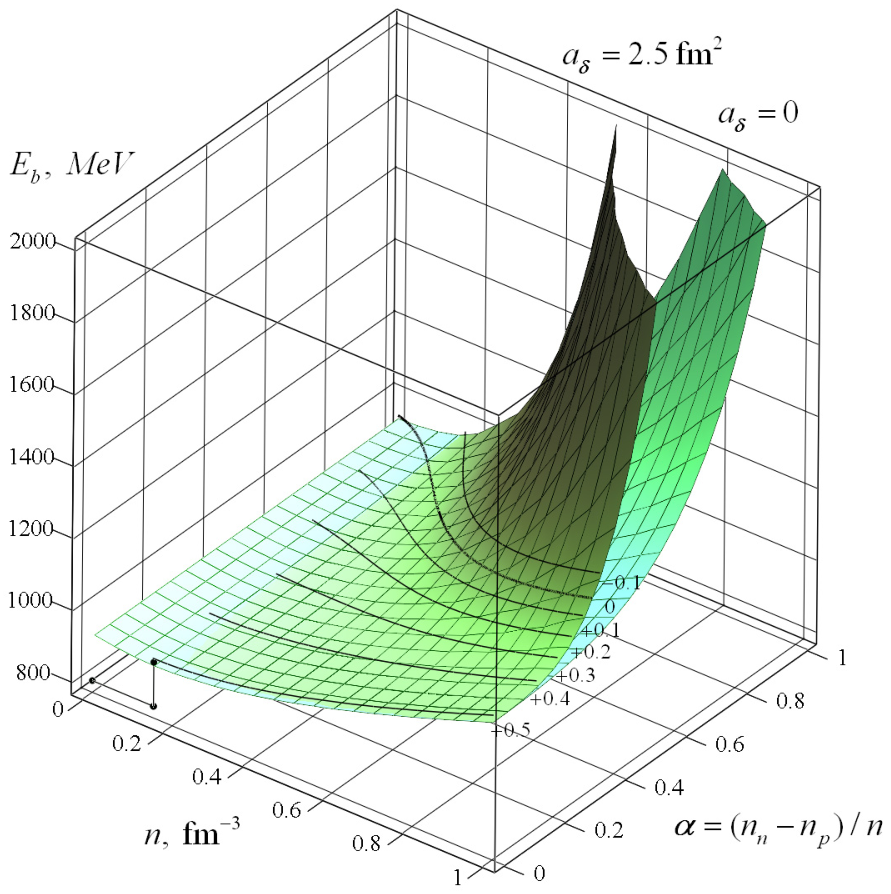,height=2.8 in}
   \caption {\small{Energy per baryon $E_{b}$ as a function of the baryon number density $n$ and
the asymmetry parameter $\alpha$ in case of a $\beta$ -equilibrium
charged $ npe$-plasma.}}
    \end{center}
  \end{minipage}\label{Fig1} \hfil\hfil
  \begin{minipage}{0.47\linewidth}
   \begin{center}
   \epsfig{file=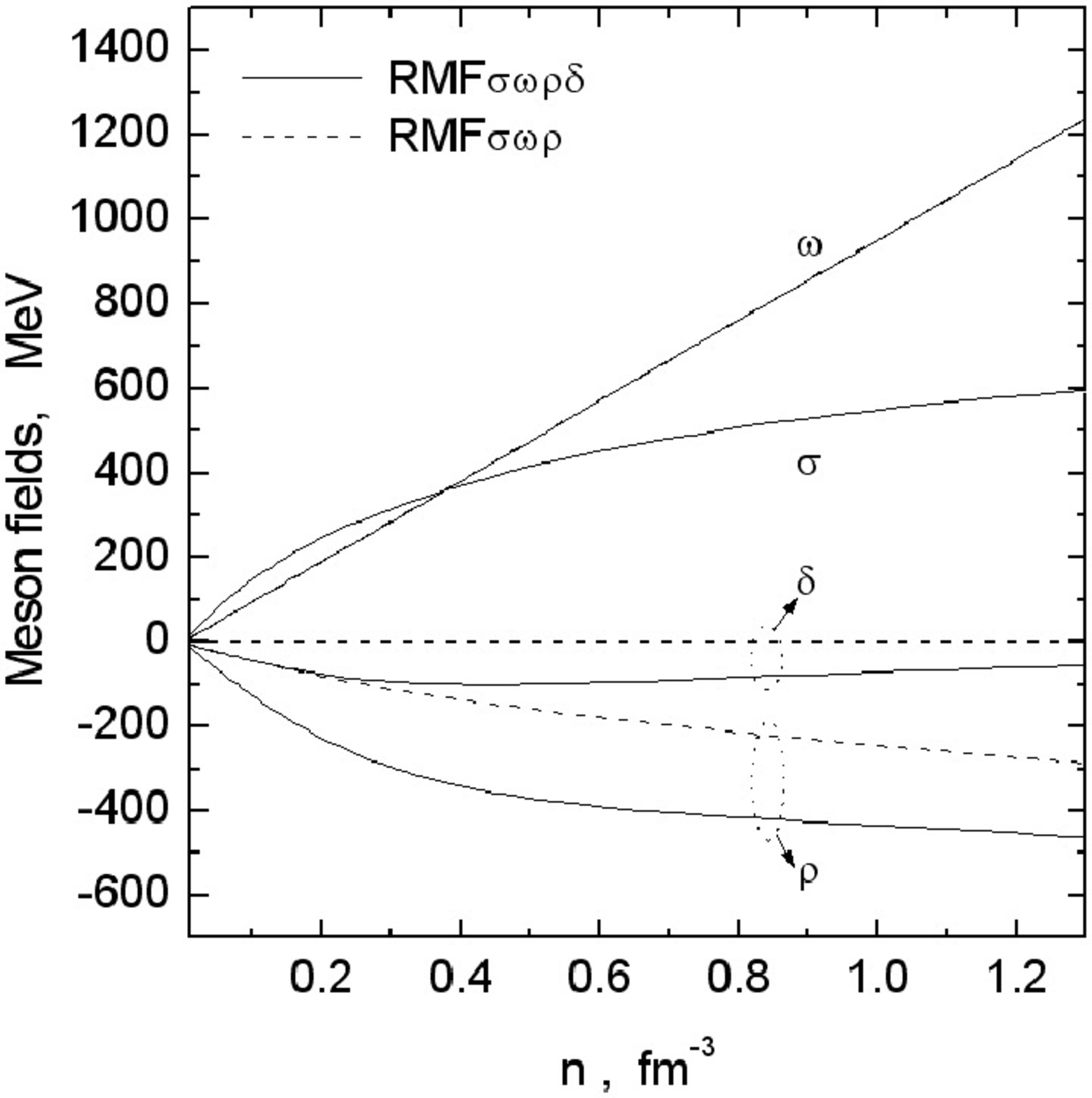,height=2.7 in}
    \caption {\small{Re-denoted meson mean-fields as a function of the baryon number density $n$ in case of a $\beta$-equilibrium charge-neutral $npe$-plasma with and without $\delta$-meson field.}}
    \end{center}
  \end{minipage}\label{Fig2}
 \end{center}
\end{figure}

In Fig 1 we illustrate the 3D-plot of the energy per baryon,
$E_{b} \left( {n,\alpha}  \right) = {{\varepsilon _{NM}}
\mathord{\left/ {\vphantom {{\varepsilon _{NM}} {n}}} \right.
\kern-\nulldelimiterspace} {n}}$, as a function of the baryon
number density $n$ and asymmetry parameter $\alpha$ in case of a
$\beta $ -equilibrium charged $npe$-plasma\cite{Al09b}. The curves
correspond to different fixed values of the charge per baryon, $q
= {{\left( {n_{p} - n_{e}} \right)} \mathord{\left/ {\vphantom
{{\left( {n_{p} - n_{e}} \right)} {n}}} \right.
\kern-\nulldelimiterspace} {n}} = {{\left( {1 - \alpha} \right)}
\mathord{\left/ {\vphantom {{\left( {1 - \alpha} \right)} {2}}}
\right. \kern-\nulldelimiterspace} {2}} - {{n_{e}} \mathord{\left/
{\vphantom {{n_{e}}  {n}}} \right. \kern-\nulldelimiterspace}
{n}}$. The thick one corresponds to $\beta$-equilibrium
charge-neutral $npe$-matter. The lower and upper surfaces
correspond to the "$\sigma \omega \rho $" and "$\sigma \omega \rho
\delta $" models, respectively. Clearly, including a $\delta$
-meson field increases the energy per baryon, and this change is
greater for larger values of the asymmetry parameter. For a fixed
value of the specific charge, the asymmetry parameter falls off
monotonically with the increase of density.

In Fig 2 we plotted the effective mean-fields of exchanged mesons,
$\sigma$, $\omega$, $\rho$ and $\delta $ as a function of the
baryon number density $n$ for the charge-neutral
$\beta$-equilibrium $npe$-plasma. The solid and dashed lines
correspond to the "$\sigma \omega \rho \delta $" and "$\sigma
\omega \rho $" models, respectively. From Fig 2 one can see that
the inclusion of the scalar-isovector virtual $\delta(a_{0}(980))$
meson results in significant changes of $\rho$ and $\delta$ meson
fields. This can result in changes of deconfinement phase
transition parameters and, thus, alter the structural
characteristics of neutron stars.

The results of our analysis show that the scalar - isovector
$\delta$-meson field inclusion leads to the increase of the EOS
stiffness of nuclear matter due to the splitting of proton and
neutron effective masses, and also the increase of asymmetry
energy (for details see Ref.\cite{Al09b}).

\subsection{Quark matter}
To describe  the quark phase an improved version of the MIT bag
model\cite{Chod74} is used, in which the interactions between
$u,~d,~s$ quarks inside the bag are taken into account in the
one-gluon exchange approximation \cite{FarJaf84}. The quark phase
consists of three quark flavors $u,~d,~s$ and electrons in
equilibrium with respect to week interactions. We choose $m_{u} =
5$ MeV, $m_{d} = 7$ MeV and $m_{s} = 150$ MeV for quark masses and
$\alpha_{s}=0.5$ for the strong interaction constant.

\subsection{Deconfinement phase transition parameters}

There are two independent conserved charges in hadron-quark phase
transition: baryonic charge and electric charge. The constituents
chemical potentials of the $npe$-plasma in $\beta$-equilibrium are
expressed through two potentials, $\mu_{b}^{(NM)}$ and
$\mu_{el}^{(NM)}$, according to conserved charges, as follows
\begin{equation}\label{chempot}
\mu_{n}=\mu_{b}^{(NM)},~~~~\mu_{p}=\mu_{b}^{(NM)}-\mu_{el}^{(NM)},
~~~~\mu_{e}=\mu_{el}^{(NM)}.
\end{equation}
In this case, the pressure $P$, energy density $\varepsilon$ and
baryon number density $n$, are functions of potentials,
 $\mu_{b}^{(NM)}$ and $\mu_{el}^{(NM)}$:
~$P_{NM}(\mu_{b}^{(NM)},\mu_{el}^{(NM)})$,
~$\varepsilon_{NM}(\mu_{b}^{(NM)},\mu_{el}^{(NM)})$,
~$n_{NM}(\mu_{b}^{(NM)},\mu_{el}^{(NM)})$.

The particle species chemical potentials for $udse$-plasma in
$\beta$-equilibrium  are expressed through the chemical potentials
$\mu_{b}^{(QM)}$ and $\mu_{el}^{(QM)}$ as follows
\begin{eqnarray}{}
\mu_{u}&=&\frac{1}{3}\left(\mu_{b}^{(QM)}-2~\mu_{el}^{(QM)}\right),\nonumber\\
\mu_{d}&=&\mu_{s}=\frac{1}{3}\left(\mu_{b}^{(QM)}+\mu_{el}^{(QM)}\right),\\
\mu_{e}&=&\mu_{d}-\mu_{u}=\mu_{el}^{(QM)}.\nonumber
\end{eqnarray}
In this case, the thermodynamic characteristics  are functions of
chemical potentials $\mu_{b}^{(QM)}$ and $\mu_{el}^{(QM)}$:
$P_{QM}(\mu_{b}^{(QM)},~\mu_{el}^{(QM)})$,~~
$\varepsilon_{QM}(\mu_{b}^{(QM)},~\mu_{el}^{(QM)})$,~~
$n_{QM}(\mu_{b}^{(QM)},~\mu_{el}^{(QM)})$.

The mechanical and chemical equilibrium conditions (Gibbs
conditions) for mixed phase are
\begin{equation}{}
\mu_{b}^{(QM)}=\mu_{b}^{(NM)}=\mu_{b},~~~~\mu_{el}^{(QM)}=\mu_{el}^{(NM)}=\mu_{el},
\end{equation}
\begin{equation}{}
P_{QM}(\mu_{b},~\mu_{el})=P_{NM}(\mu_{b},~\mu_{el}).
\end{equation}

The volume fraction of quark phase is
\begin{equation}
\chi=V_{QM}/\left(V_{QM}+V_{NM}\right),
\end{equation}
\noindent where $V_{QM}$ and $V_{NM}$ are volumes occupied by
quark matter and nucleonic matter, respectively.

We applied the global electrical neutrality condition for mixed
quark-nucleonic matter, according to Glendenning \cite{Gl92,
Gl00},
\begin{eqnarray}{}
(1-\chi)\left[n_{p}(\mu_{b},\mu_{el})-n_{e}(\mu_{el})\right]\nonumber\\+\chi\left[\frac{2}{3}~n_{u}(\mu_{b},\mu_{el})-\frac{1}{3}~n_{d}(\mu_{b},\mu_{el})-\frac{1}{3}~n_{s}(\mu_{b},\mu_{el})-n_{e}(\mu_{el})\right]=0.
\end{eqnarray}
The baryon number density in the mixed phase is determined as
\begin{eqnarray}{}
n=(1-\chi)\left[n_{p}(\mu_{b},\mu_{el})+n_{n}(\mu_{b},\mu_{el})\right]\nonumber\\+\frac{1}{3}~\chi\left[n_{u}(\mu_{b},\mu_{el})+n_{d}(\mu_{b},\mu_{el})+n_{s}(\mu_{b},\mu_{el})\right],
\end{eqnarray}
and the energy density is
\begin{eqnarray}{}
\varepsilon=(1-\chi)\left[\varepsilon_{p}(\mu_{b},\mu_{el})+\varepsilon_{n}(\mu_{b},\mu_{el})\right]\nonumber\\+\chi\left[\varepsilon_{u}(\mu_{b},\mu_{el})+\varepsilon_{d}(\mu_{b},\mu_{el})+\varepsilon_{s}(\mu_{b},\mu_{el})\right]+\varepsilon_{e}(\mu_{el}).
\end{eqnarray}

In case of $\chi=0$, the chemical potentials $\mu_{b}^{N}$ and
$\mu_{el}^{N}$ corresponding to the lower threshold of a mixed
phase, are determined by solving Eqs. (12) and (14). This allows
to determine the lower boundary parameters $P_{N}$,
$\varepsilon_{N}$ and $n_{N}$. Similarly, we determine the upper
boundary values of mixed phase parameters, $P_{Q}$,
$\varepsilon_{Q}$ and $n_{Q}$, for $\chi=1$. The system of Eqs.
(12), (14),(15) and (16) makes it possible to determine EOS of
mixed phase between this critical states.

\begin{table}[ht]
\begin{center}
\begin{tabular}{l|c|c|c|c|c|c}  Model& $n_{N}$ & $n_{Q}$  & $P_{N}$ &
$P_{Q}$ & $\varepsilon_{N}$ & $\varepsilon_{Q}$ \\
 & (fm$^{-3})$ & (fm$^{-3}$) & (MeV/fm$^{3}$) & (MeV/fm$^{3}$) &
(MeV/fm$^{3}$) & (MeV/fm$^{3}$) \\
     \hline B60$\sigma \omega \rho $ & 0.0717 & 1.083 &  0.336 & 327.747 &  67.728 & 1280.889 \\
  \hline B60$\sigma\omega\rho\delta$ & 0.0771 & 1.083 &  0.434 & 327.745 &  72.793 & 1280.884 \\
    \hline B100$\sigma \omega \rho $ & 0.2596 & 1.436 & 18.025 & 471.310 & 253.814 & 1870.769 \\
 \hline B100$\sigma\omega\rho\delta$ & 0.2409 & 1.448 & 16.911 & 474.368 & 235.029 & 1889.336 \\
\end{tabular}
\caption{The Mixed phase threshold parameters without ($\sigma
\omega \rho $) and with ($\sigma \omega \rho \delta$) a $\delta$
-meson field for bag parameter values, $B=60$ MeV/fm$^3$ and
$B=100$ MeV/fm$^3$.} \label{T-2}
\end{center}
\end{table}

Note, that in the case of an ordinary first-order phase transition
both nuclear and quark matter are assumed to be separately
electrically neutral, and at some pressure $P_{0}$, corresponding
to the coexistence of the two phases, their baryon chemical
potentials are equal, i.e.,
\begin{equation}
\label{eq45} \mu _{NM} \left( {P_{0}}  \right) = \mu _{QM} \left(
{P_{0}}  \right).
\end{equation}
Such phase transition scenario is known as phase transition with
constant pressure (Maxwell construction).

\begin{figure}[ht]
\begin{center}
  \begin{minipage}{0.47\linewidth}
   \begin{center}
   \epsfig{file=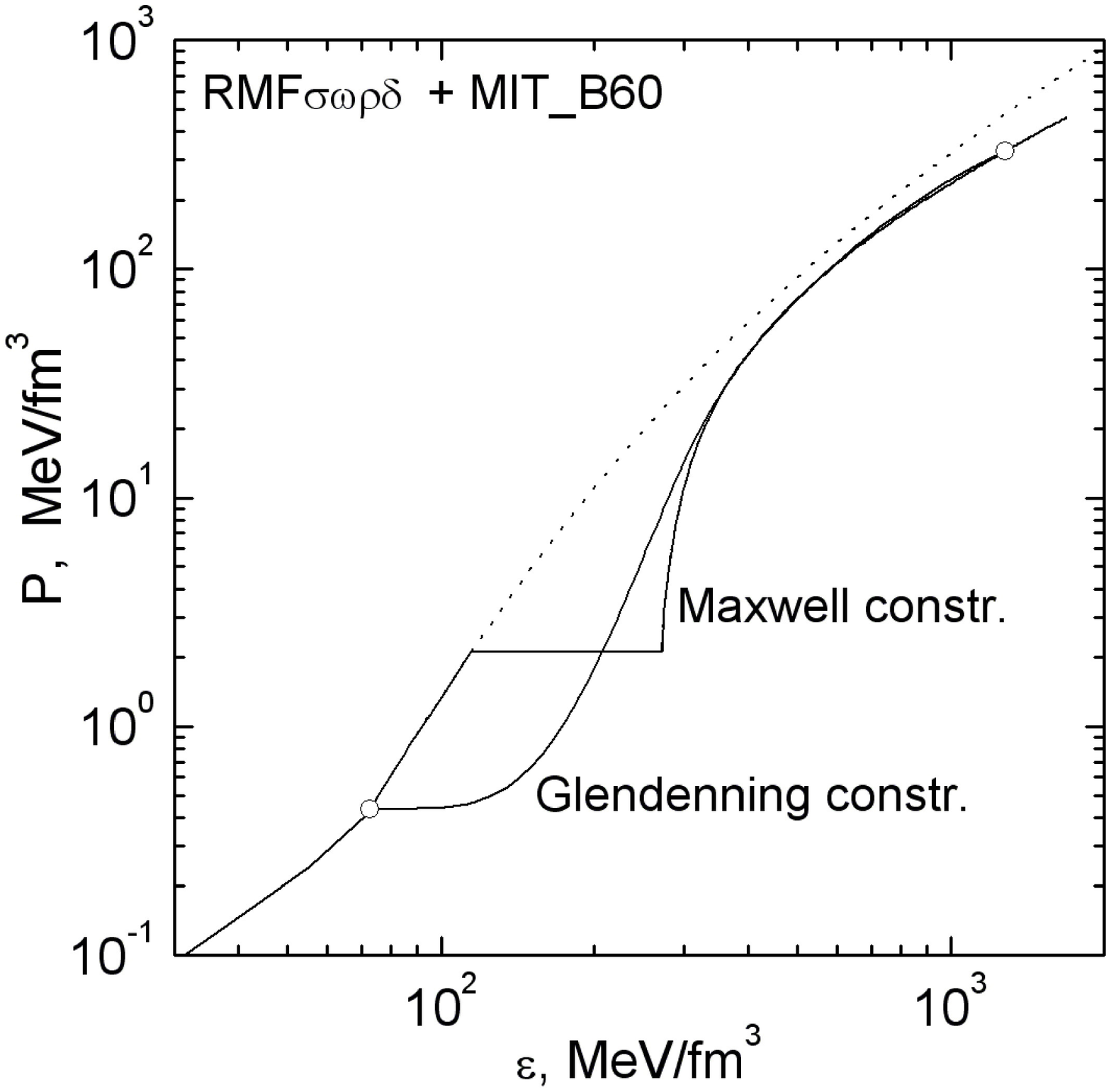,height=2.8 in}
   \caption {\small{EOS of neutron star matter with the deconfinement
   phase transition for a bag constant $B=60$ MeV/fm$^3$. For
   comparison we plot both the Glendenning and Maxwell
   constructions. Open circles represent the mixed phase boundaries.}}
    \end{center}
  \end{minipage}\label{Fig3} \hfil\hfil
  \begin{minipage}{0.47\linewidth}
   \begin{center}
   \epsfig{file=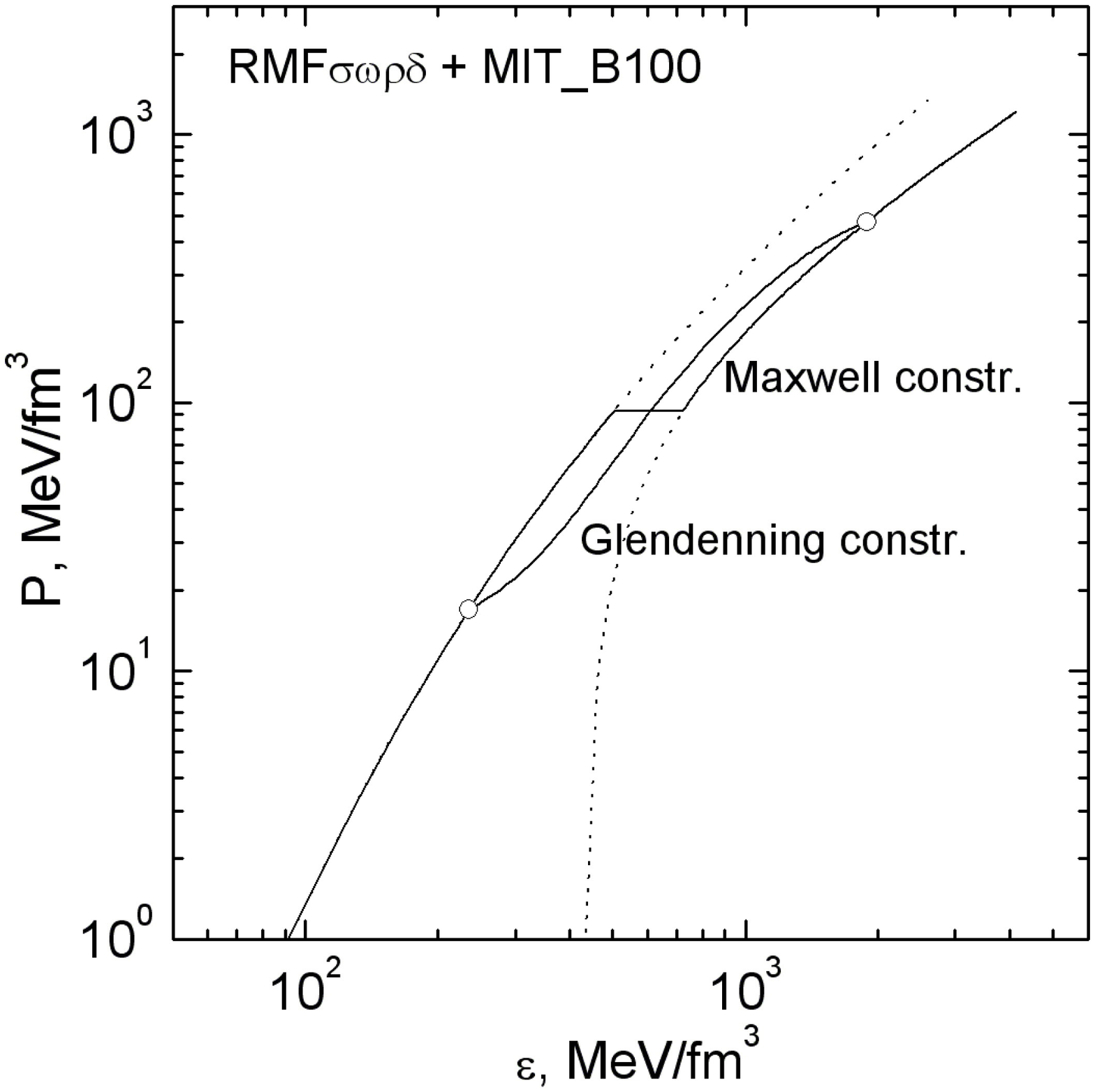,height=2.8 in}
   \caption {\small{Same as in Fig 3, but for $B=100$ MeV/fm$^3$.} \bigskip\bigskip\bigskip\bigskip\bigskip}
    \end{center}
  \end{minipage}\label{Fig4}
 \end{center}
\end{figure}
Table 2 represents the parameter sets of the mixed phase both with
and without $\delta$-meson field. It is shown that the presence of
$\delta$-field alters threshold characteristics of the mixed
phase. For $B=60$ MeV/fm$^3$ the lower threshold parameters,
$n_{N}$, $\varepsilon_{N}$, $P_{N}$ are increased, meanwhile the
upper ones $n_{Q}$, $\varepsilon_{Q}$, $P_{Q}$ are slowly
decreased. For $B=100$ MeV/fm$^3$ this behavior changes to
opposite.

In Fig 3 and Fig 4 we plot the EOS of star matter with
deconfinement phase transition for two values of bag constant,
$B=60$ MeV/fm$^3$ and $B=100$ MeV/fm$^3$, respectively. The dotted
curves correspond to pure nucleonic and quark matters without any
phase transition, while the solid lines correspond to two
alternative phase transition scenarios. Open circles show the
boundary points of the mixed phase.
\begin{figure}[ht]
\begin{center}
  \begin{minipage}{0.47\linewidth}
   \begin{center}
   \epsfig{file=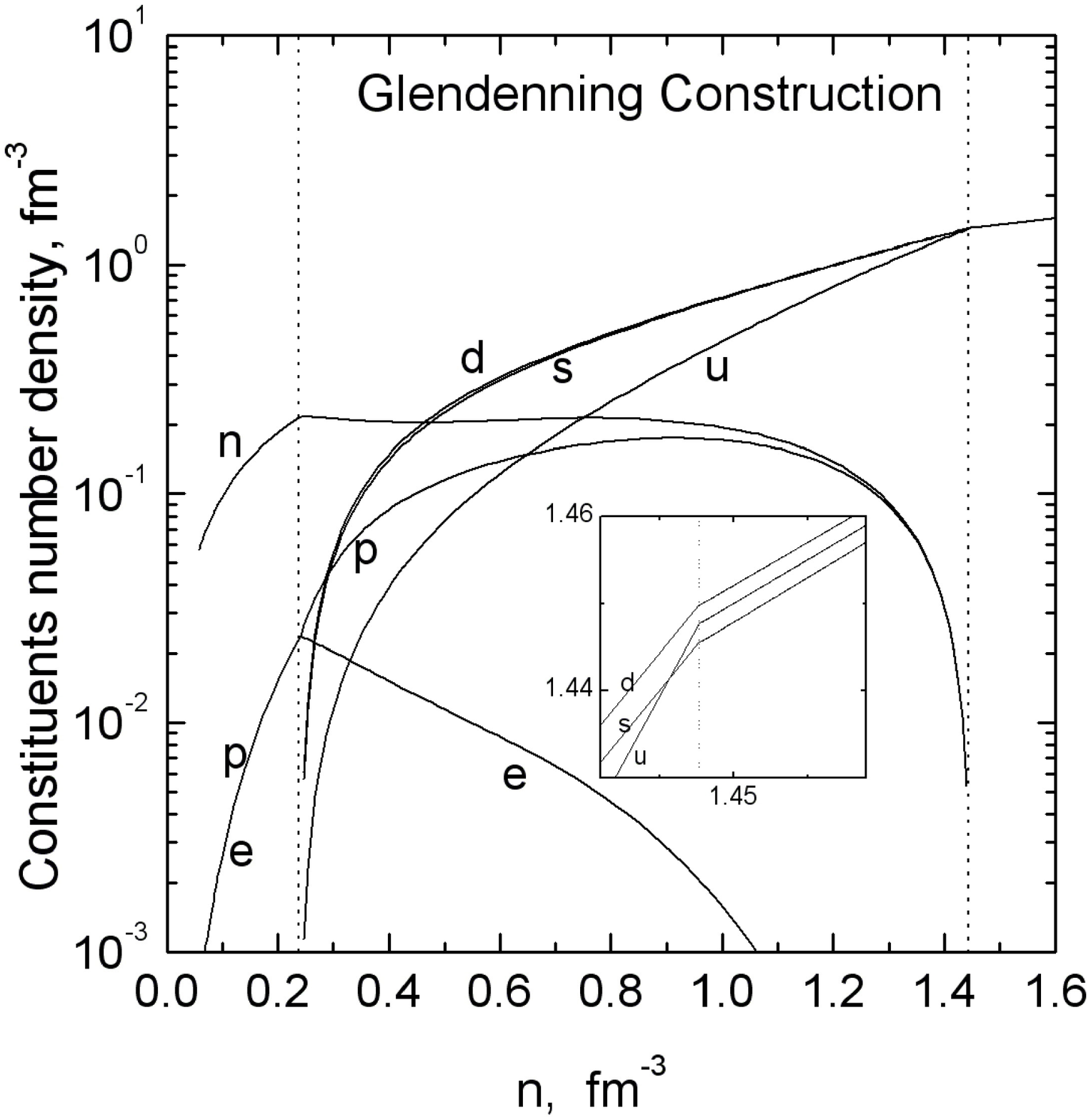,height=2.7 in}
   \caption {\small{Constituents number density versus baryon number density $n$ for $B=100$ MeV/fm$^3$ in case of Glendenning construction. Vertical dotted lines represent the mixed phase boundaries.}}
    \end{center}
  \end{minipage}\label{Fig5} \hfil\hfil
  \begin{minipage}{0.47\linewidth}
   \begin{center}
   \epsfig{file=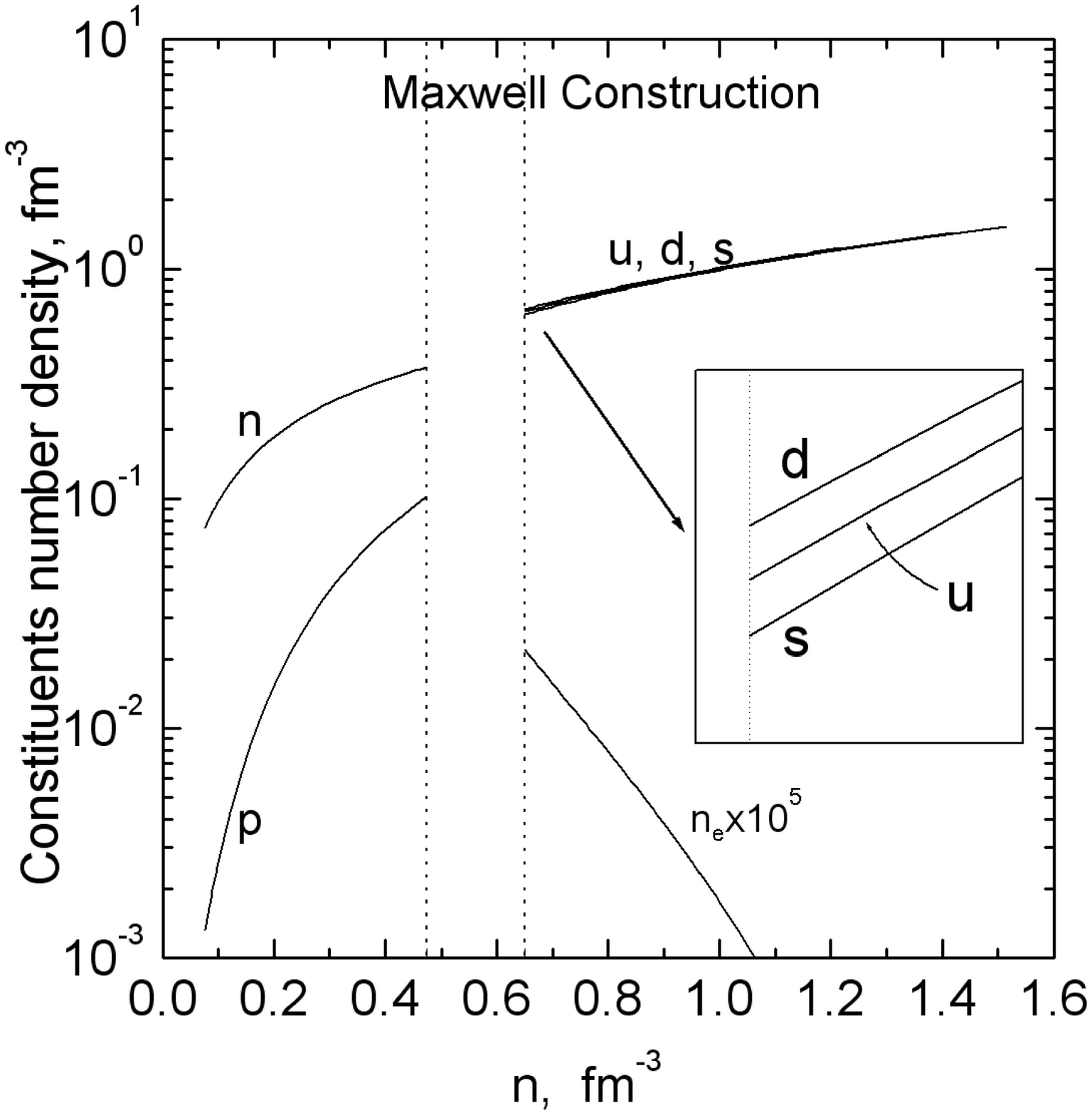,height=2.8 in}
   \caption {\small{Same as in Fig 5, but for Maxwell construction. Vertical dotted lines represent the density jump boundaries. }\bigskip\bigskip}
    \end{center}
  \end{minipage}\label{Fig6}
 \end{center}
\end{figure}

In Fig 5 we plot the particle species number densities as a
function of baryon density $n$ for Glendenning construction.
Quarks appear at the critical density $n_{N}=0.241$ fm$^{-3}$. The
hadronic matter completely disappears at $n_{Q}=1.448$ fm$^{-3}$,
where the pure quark phase occurs. Fig 6 show the constituents
number density as a function of baryon number density $n$ for
$B=100$ MeV/fm$^3$, when phase transition described according to
Maxwell construction. Maxwell construction leads to the appearance
of a discontinuity. In this case, the charge neutral nucleonic
matter at baryon density $n_{1}=0.475$ fm$^{-3}$ coexisted with
the charge neutral quark matter at baryon density $n_{2}=0.650$
fm$^{-3}$. Thus, the density range $n_{1}<n<n_{2}$ is forbidden.
In the Maxwell construction case the chemical potential of
electrons, $\mu_{e}$, has a jump at the coexistence pressure
$P_{0}$. Notice, that such discontinuity behaviour takes place
only in usual first-order phase transition, i.e., in the Maxwell
construction case.

\section{Properties of hybrid stars}

Using the EOS obtained in previous section, we calculate the
integral and structure characteristics of Neutron stars with quark
degrees of freedom.

The hydrostatic equilibrium properties of spherical symmetric and
isotropic compact stars in general relativity is described by the
Tolman-Oppenheimer-Volkoff(TOV) equations\cite{TOV39}:
\begin{eqnarray}\label{TOV}
\frac{dP}{dr}&=&-\frac{G}{r^{2}}\frac{(P+\varepsilon)(m+4\pi
r^{3}P)}{1-2G~m/r},\\
\frac{dm}{dr}&=&4\pi r^2 \varepsilon,
\end{eqnarray}
\noindent where $G$ is the gravitational constant, $r$ is the
distance from the center of star, $P(r)$, $\varepsilon(r)$ are the
pressure and energy density at the radius $r$, respectively, and
$m(r)$ is the mass inside a sphere of radius $r$. For integration
of the TOV equations it is necessary to know the EOS of neutron
star matter in a form $\varepsilon(P)$. Using the neutron star
matter EOS , obtained in previous section, we have integrated the
Tolman-Oppenheimer Volkoff equations and obtained the
gravitational mass $M$ and the radius $R$ of compact stars (with
and without quark degrees of freedom) for the different values of
central pressure $P_{c}$.
\begin{figure}[ht]
\begin{center}
  \begin{minipage}{0.47\linewidth}
   \begin{center}
   \epsfig{file=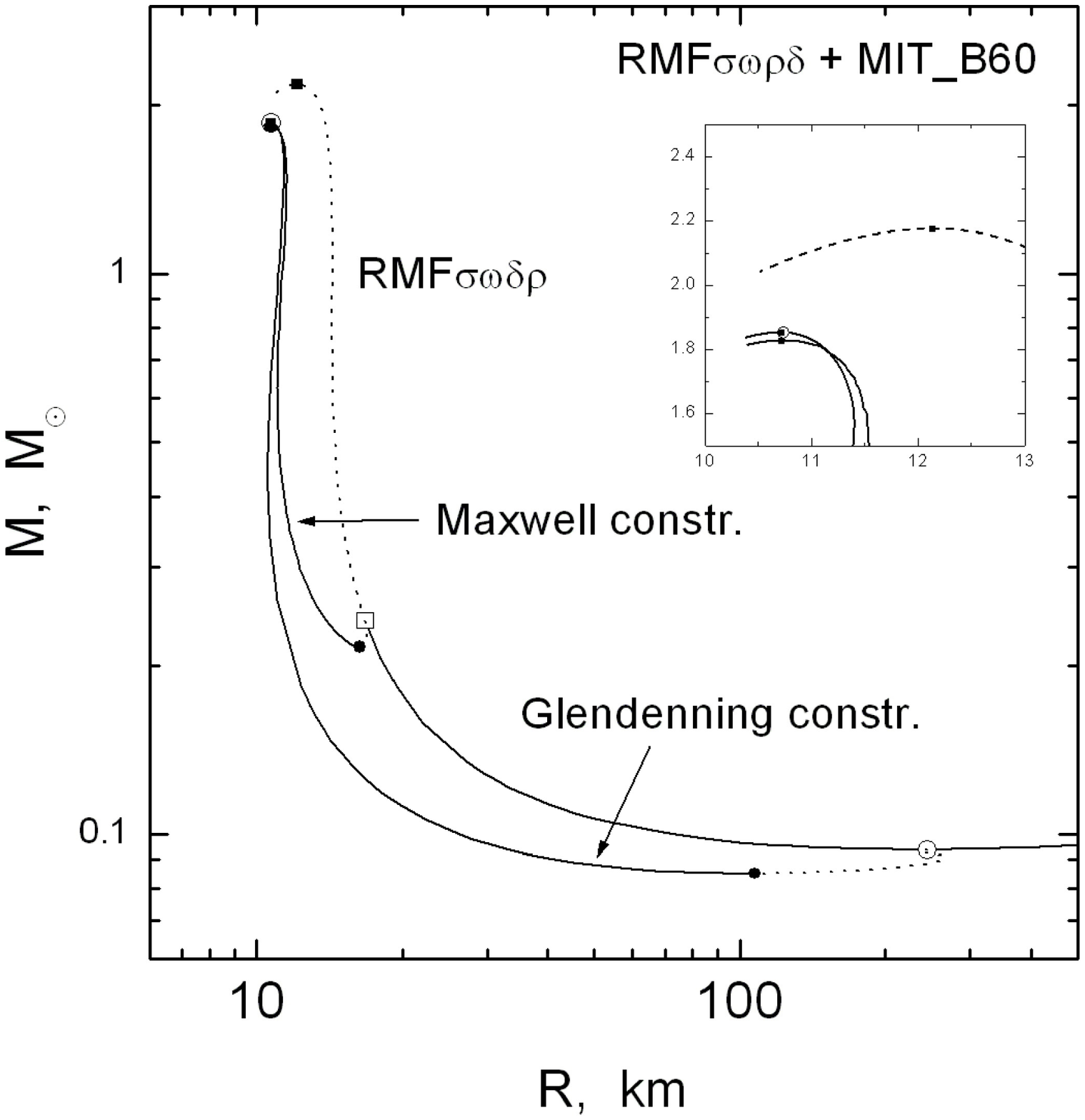,height=2.8 in}
   \caption {\small{The mass-radius relation of neutron star with different deconfinement
   phase transition scenarios for bag constant $B=60$ MeV/fm$^3$. Open circles and squares denote the critical configurations for Glendenning and Maxwellian type transitions, respectively. Solid circles and squares denote hybrid stars with minimal and maximal masses, respectively.   . }}
    \end{center}
  \end{minipage}\label{Fig7} \hfil\hfil
  \begin{minipage}{0.47\linewidth}
   \begin{center}
   \epsfig{file=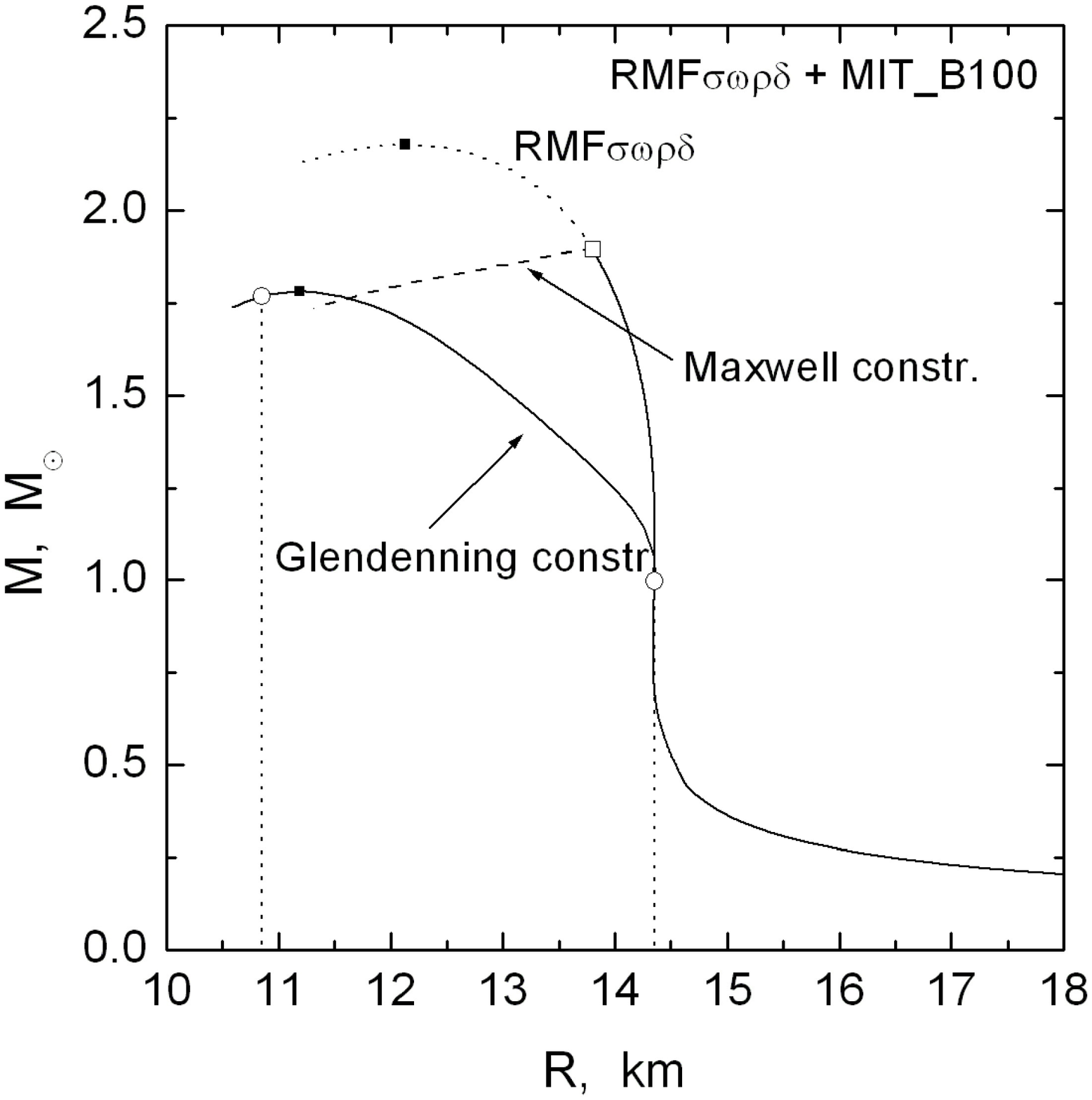,height=2.8 in}
   \caption {\small{Same as in Fig 7, but for $B=100$ MeV/fm$^3$. }\bigskip\bigskip\bigskip\bigskip\bigskip\bigskip\bigskip\bigskip}
    \end{center}
  \end{minipage}\label{Fig8}
 \end{center}
\end{figure}

Fig 7 and Fig 8 illustrates the $M(R)$ dependence of neutron stars
for two values of bag constant $B=60$ MeV/fm$^3$ and $B=100$
MeV/fm$^3$, respectively. We can see, that the behaviour of
mass-radius dependence significantly differs for two types of
phase transitions. Fig 7 shows, that for $B=60$ MeV/fm$^3$ there
are unstable region, where $dM/dP_{c}<0$, between two stable
branches of compact stars, corresponding to configurations with
and without quark matter. In this case, there is a nonzero minimum
value of the radius of quark phase core. Accretion of matter on a
critical neutron star configuration will then result in a
catastrophic rearrangement of the star, forming a star with a
quark matter core. The range of mass values for stars, containing
the mixed phase, is $[0.085 M_{\odot}; 1.853M_{\odot}]$ for $B=60$
MeV/fm$^3$, and is $[0.997M_{\odot}; 1.780M_{\odot}]$ for $B=100$
MeV/fm$^3$. In case of Maxwellian type phase transition the
analogous range is $[0.216M_{\odot}; 1.828M_{\odot}]$ for $B=60$
MeV/fm$^3$. From Fig 8 one can observe that when $B=100
$MeV/fm$^3$ MeV/fm$^3$, the star configurations with deconfined
quark matter is unstable. Thus, the stable neutron star maximum
mass is $1.894M_{\odot}$. Our analysis show, that the pressure
upper threshold value for mixed phase is larger, then the
pressure, corresponding to maximum mass configuration. This means,
that in the model, considered here, the mixed phase can exist in
the center of compact stars, but no pure quark matter can exist.

\section{Conclusion}

In this paper we have studied the deconfinement phase transition
of neutron star matter, when the nuclear matter is described in
the RMF theory with $\delta$-meson effective fields. We show that
the inclusion of scalar – isovector $\delta$-meson field terms
leads to the stiff nuclear matter EOS. The presence of scalar –
isovector $\delta$-meson field alters the threshold
characteristics of the mixed phase. For $B=60$ MeV/fm$^3$, the
lower threshold parameters, $n_{N}$, $\varepsilon_{N}$, $P_{N}$
are increased, meanwhile the upper ones, $n_{Q}$,
$\varepsilon_{Q}$, $P_{Q}$ are slowly decreased. For $B=100$
MeV/fm$^3$ this behavior changes to opposite.

For EOS used in this study, the central pressure of the maximum
mass neutron stars is less than the mixed phase upper threshold
$P_Q$. Thus, the corresponding hybrid stars do not contain pure
strange quark matter core.

\bigskip
The author would like to thank G.S.Hajyan and Yu.L.Vartanyan for
useful discussions on issues related to the subject of this
research.


\begin{thebibliography}{99}

\bibitem{Wal74} J.D. Walecka, Ann. Phys., \textbf{83}, 491 (1974).
\bibitem{SW86} B.D. Serot and J.D. Walecka, in: Adv. in Nucl. Phys., eds. J. W. Negele and E. Vogt, vol. \textbf{16} (1986).
\bibitem{SW97} B.D. Serot, J.D. Walecka, Int.J.Mod.Phys. \textbf{E6}, 515 (1997).
\bibitem{Lal97} G.A. Lalazissis, J. Konig, and P. Ring, Phys.Rev. \textbf{C55}, 540 (1997).
\bibitem{Typel99} S. Typel, H.H. Wolter, Nucl. Phys. \textbf{A656}, 331 (1999).
\bibitem{Ko_Li96} C.M. Ko, G.Q. Li, Journal of Phys., \textbf{G22},1673 (1996).
\bibitem{PR_Lal07} V. Prassa, G. Ferini, T. Gaitanos, H.H. Wolter, G.A. Lalazissis, and
M. Di Toro, arXiv:0704.0554 v1 [nucl-th] (2007)
\bibitem{Mill95} H. Miller, B.D. Serot, Phys. Rev. \textbf{C52}, 2072 (1995).
\bibitem{Kubis97} S. Kubis, M. Kutschera, Phys. Lett., \textbf{B399},191 (1997).
\bibitem{Liu02} B. Liu, V. Greco, V. Baran, M. Colonna, M. Di Toro, Phys.Rev.
\textbf{C65}, 045201 (2002).
\bibitem{Greco03} V. Greco, M. Colonna, M. Di Toro, F. Matera, Phys. Rev. \textbf{C67},
015203 (2003).
\bibitem{Gl92} N.K. Glendenning, Phys. Rev. \textbf{D 46}, 1274 (1992).
\bibitem{Heis93} H. Heiselberg, C.J. Pethick, and E.S. Staubo, Phys. Rev. Lett. \textbf{70}, 1355 (1993). H. Heiselberg and M. Hjorth- Jensen, arXiv: 9902033 v1, [nucl-th] (1999).
\bibitem{Al09a} G.B. Alaverdyan, Astrophysics \textbf{52}, 132 (2009).
\bibitem{BBP71} G. Baym, H. Bethe, Ch. Pethick, Nucl.Phys.,\textbf{ A175}, 255, 1971.
\bibitem{Gl00} N. K. Glendenning, Compact Stars, Springer (2000).
\bibitem{Al09b} G. B. Alaverdyan, Gravitation $\&$ Cosmology \textbf{15}, 5 (2009).
\bibitem{Chod74} A. Chodos, R. L. Jaffe, K. Johnson, C. B. Thorn, V. F. Weisskopf,
Phys.Rev.\textbf{D9}, 3471 (1974).
\bibitem{FarJaf84} E. Farhi, R. L. Jaffe, Phys.Rev. \textbf{D30}, 2379 (1984).
\bibitem{TOV39} R. Tolman, Phys.Rev. 55, 364, 1939; J. Oppenheimer and G. Volkoff,
Phys.Rev. \textbf{55}, 374 (1939).

\end{thebibliography}
\end{document}